\documentstyle[aps,prb,preprint]{revtex}
\begin{document}
\draft \title{Landauer Theory, Inelastic Scattering and
Electron Transport in Molecular Wires}
\author{Eldon G. Emberly and George Kirczenow}
\address{Department
of Physics, Simon Fraser University, Burnaby, B.C., Canada V5A
1S6}
\date{\today}
\maketitle
\begin{abstract}
In this paper we address the topic of inelastic electron
scattering in mesoscopic quantum transport. For systems where
only elastic scattering is present, Landauer theory provides
an adequate description of transport that relates the
electronic current to single-particle transmission and
reflection probabilities. A   formalism proposed recently by
Bon$\breve{c}$a and Trugman facilitates the calculation of the
one-electron transmission and reflection probabilities for
{\em inelastic} processes in mesoscopic conductors connected to
one-dimensional ideal leads. Building on their
work, we have developed a self-consistent procedure for the
evaluation of the non-equilibrium electron distributions in
ideal leads connecting such mesoscopic conductors to
electron reservoirs at finite temperatures and voltages.
We evaluate the net electronic current flowing through the
mesoscopic device by utilizing these non-equilibrium
distributions. Our approach is a generalization of Landauer
theory that takes account of the Pauli exclusion principle for
the various competing elastic and inelastic processes while
satisfying the requirement of particle conservation.   As an
application we examine the influence of elastic and inelastic
scattering on conduction through a two site molecular wire with
longitudinal phonons using the Su-Schrieffer-Heeger model of
electron-phonon coupling.
\end{abstract}
\section{Introduction}
The role of inelastic scattering in electron transport through
mesoscopic systems is a topic of current theoretical and
experimental interest. Recent experiments have revealed the
importance of the electron-phonon interaction in transport
measurements performed on certain mesoscopic systems. In
particular inelastic scattering effects have been observed
directly in STM measurements of the differential conductance of
molecules adsorbed on metallic
substrates.\cite{Stip1,Stip2,Stip3} Theoretically the effects
of inelastic scattering on electron transport through
mesoscopic semiconductor devices have been investigated by a
variety of methods ranging from Green's function
techniques\cite{Win88,Gelf89,Stov91} to the Fermi golden
rule.\cite{Turl91} Some theoretical models have also been
proposed to elucidate the effects of molecular vibrations on
electron tunneling through molecules contacted by an
STM.\cite{Pers87,Bin85,Gata93} Recent work has provided a
method for calculating the electronic current at finite
temperatures for electrons tunneling through a one-dimensional
conductor in the presence of phonons.\cite{Hau98} These
approaches have yielded valuable insights into the behavior of
specific systems.

In the absence of inelastic scattering and electron
correlations, Landauer theory\cite{Lan57} provides a general
framework for calculations of the electronic current through
mesoscopic conductors that are coupled to ideal single- or
multi-channel quasi-one-dimensional leads.  It relates the
electronic current to the transmission probability for an
electron incident from the source lead to scatter elastically
through the conductor and into the drain. The transmission
probability is found by solving the single-electron quantum
scattering problem.

When inelastic processes such as phonon emission and absorption
are considered, electron scattering becomes a many-body
phenomenon involving electrons and the various excited phonon
states of the system.  A non-perturbative approach has been
proposed by Bon$\breve{c}$a and Trugman for treating this
scattering problem for a system consisting of a mesoscopic
conductor (which supports the phonon modes) coupled to two
single-channel ideal leads that act as the electron source and
drain.\cite{Bonc95} Their method approximates the many-body
problem by a multi-channel single-electron scattering problem
which can be solved exactly.  Here each channel corresponds to
a different vibrational quantum state of the mesoscopic
conductor. Using their approach it is possible to determine the
transmission and reflection probabilities for all inelastic and
elastic scattering events that an electron may suffer in the
conductor as it goes from the source to the drain. This method
has been applied to study the effects of phonons on electron
transmission through one-dimensional conductors where the
electron-phonon coupling has been modelled using the Holstein
Hamiltonian\cite{Hol59,Bonc97,Hau98} and the
Su-Schrieffer-Heeger (SSH) Hamiltonian\cite{Su79,Ness99}.

The present article is complementary to the above theoretical
work. Its purpose is to generalize the Landauer theory of
electrical conduction to mesoscopic systems with
electron-phonon scattering, assuming that the latter is
described accurately within the multi-channel single-electron
scattering approach of Bon$\breve{c}$a and
Trugman\cite{Bonc95}. I.e., given the
Bon$\breve{c}$a-Trugman\cite{Bonc95} solution of the inelastic
scattering problem, our objective is to calculate the electric
current by appropriately generalizing Landauer theory.

When transport in a many-electron system is described in terms
of one-electron scattering processes, the Pauli exclusion
principle needs to be considered since in general electrons
involved in different transitions may compete to occupy the
same final state. In treatments of transport based on the
Boltzmann equation, if the electron scattering is elastic, this
competition has no effect because of a mutual cancellation of
terms.\cite{Peierls1} However such a cancellation does not
occur when scattering is inelastic.\cite{Peierls2} In Landauer
theory scattering is assumed to be elastic and the electronic
current is carried by electrons which occupy single-particle
scattering states that extend across the mesoscopic conductor
from the source to the drain. These scattering states are
orthogonal to each other.\cite{Bal} Because of this
orthogonality, if an electron populates one of these scattering
states it does not compete with any of the other filled
scattering states. Thus in Landauer theory (as in the elastic
case of the Boltzmann equation) the Pauli exclusion principle
for the final states in the scattering processes does not play
a role in the determination of the electronic current.

In our generalization of Landauer theory that includes
inelastic scattering, this is no longer true: Different
scattering processes can send electrons to the same outgoing
state and the Pauli principle plays an important role in
determining the occupations of the outgoing channels.  In the
Boltzman treatment of inelastic scattering the Pauli principle
is applied to the non-equilibrium one-electron distribution
function for the system since transport is an inherently
non-equilibrium phenomenon. Here we apply an analogous
principle: In our extension of Landauer theory the Pauli
exclusion principle is applied to the non-equilibrium electron
distributions in the {\em outgoing} channels of the scattering
process and these distributions must be determined
self-consistently. Since some processes may be excluded
because of the Pauli principle, overall particle conservation is
another consideration that must be taken into account when
trying to generalize Landauer theory to include inelastic
scattering.

We have developed a practical method for determining these
non-equilibrium electron distributions for a class of
mesoscopic systems. These systems are similar to those
considered by Bon$\breve{c}$a and Trugman\cite{Bonc95} in that
they consists of a mesoscopic conductor which is attached to
two single mode ideal leads which act as a source and drain,
and phonons are assumed to be present only on the conductor and
not on the ideal leads. We consider scattering processes which
change the conductor's phonon state or leave it unchanged.
Each of these processes contributes to the non-equilibrium
electron distribution.  We determine each contribution
self-consistently (i.e., in the presence of the other processes
that can occur) in terms of the transmission and reflection
probabilities for the given process and the Fermi functions
that describe the incoming electron distributions in the left
and right leads. Included in this self-consistent method is the
constraint of particle conservation. We do not however include
cumulative effects of the current on the phonon distribution in
the conductor; i.e., we assume that the conductor is in contact
with a heat bath.  With the non-equilibrium electron
distribution determined we then evaluate the current flowing
through this system for finite temperatures and bias voltages.

As an application of the above theoretical approach we
calculate the electronic current as a function of voltage for
a two site molecule.  The molecule has two single-particle
molecular energy levels through which electrons from the leads
can resonantly tunnel.  It also has two longitudinal
phonon modes which can be excited. A SSH Hamiltonian is used
to represent the electron-phonon coupling. For our model system
one of the phonon modes is chosen to have an energy equal to
the gap between the two electronic energy levels.  Because of
this, electrons which are resonant with the higher energy
level can emit a phonon and be resonant with the lower level.
Such a process is found to be highly transmitting.  When the
voltage is high enough that there is a sufficient number
of electrons in the left lead that can undergo this process,
we find that the current in the right lead due to elastically
scattered electrons is reduced significantly due to competition
with these inelastically scattered electrons. Thus we find
that the mutual exclusion between the scattered electrons
that make up the non-equilibrium distribution in the drain lead
can have important consequences for the electronic current
flowing through the molecule.

In Sec. II we describe the class of systems that we study and
explain how the scattering states are calculated. In Sec. III
we present our generalization of the Landauer theory and our
method for calculating the non-equilibrium the electron
distribution and associated electronic currents.  We then
apply our methodology to a simple model for a two site
molecular chain and calculate its current-voltage
characteristics in Sec. IV. Our conclusions are summarized in
Sec. V.

\section{Inelastic Scattering in Electron Transport through a
Mesoscopic Conductor}
In the absence of inelastic scattering and electron
correlations, electron transport through a mesoscopic conductor
can be described in terms of the probability $T(E,V)$ that a
single electron with energy $E$ scatters through the conductor
from the source to the drain at an applied bias voltage
$V$.\cite{Lan57} When phonons (or vibrational modes) of the
conductor are considered, an electron entering from the source
can suffer inelastic collisions by absorbing or creating
phonons before entering the drain. Each of these processes can
be described by its own transmission probability. If we assume
that the source and drain are single mode leads which are free
from phonons, we can characterize the above single electron
scattering process by a transmission probability
$T^{\alpha,\alpha'}_{L\rightarrow R}(E,E')$. This describes an
electron that enters from the left lead $(L)$ with energy $E$,
suffers inelastic collisions on the conductor which change the
phonon population on the conductor from $\alpha$ to $\alpha'$,
and then scatters into the right lead $(R)$ with energy $E'$.
Similarly, an electron incident from the left lead can suffer a
collision and be reflected back into the left lead.  This is
characterized by a reflection probability
$R^{\alpha,\alpha'}_{L}(E,E')$.  Elastic scattering processes
are included with $\alpha'$ being equal to $\alpha$.  For all
processes the total energy is conserved, so $E'$ plus the net
energy of all of the created and destroyed phonons is equal to
$E$. We now describe a method for calculating these
transmission and reflection probabilities for a mesoscopic
structure.

A schematic of our model for the mesoscopic conductor, which
consists of a source (left) and a drain (right) lead attached
to a one-dimensional conductor/wire is depicted in
Fig.~\ref{fig1}.  We assume that the leads are ideal and that
they only have one electronic channel. Phonons are considered
to be only on the conductor and not on the ideal leads. An
electron flows through the wire under an applied bias $V$. We
assume that the voltage drop only occurs at the interface
between the wire and the leads with the left lead held at
$-V/2$ and the right lead held at $+V/2$. We solve the
transport problem by solving Schr\"{o}dinger's equation for the
many-body wavefunction of the electron-phonon system from which
we extract the transmission and reflection coefficients
mentioned above.  These will be used in the calculation of the
non-equilibrium electron distributions and the current which
follows in the next section.

The Hamiltonian $H$ of this system can be split up into the
Hamiltonians for the isolated systems consisting of the left
lead $H_L$, the right lead $H_R$, and the conductor $H_C$.  It
also includes the coupling between the wire and the left
and right leads, $V_L$ and $V_R$ respectively. Thus we write,
\begin{equation}
H =  H_L + H_R + H_C + V_L + V_R
\end{equation}
We treat each of the above Hamiltonians in the tight binding
approximation. We model the wire using a set of orthogonal
atomic orbitals $\{|i \rangle \}$.  Each atomic state has an
energy $e_i$ with a corresponding creation operator $b_{i}^+$
which creates an electron in that state. In our Hamiltonian for
the wire an electron can hop from orbital $i$ to all nearest
neighbor orbitals $j$, and this is governed by the hopping
parameter $t_{i,j}^0$ for the equilibrium configuration of
atoms. The wire also has a discrete set of vibrational modes,
each characterized by a frequency $\omega_k$. The phonons on
the wire are described by the state $|\alpha\rangle =
|\{n_k\}\rangle$, where $\{n_k\}$ is the set of mode occupation
numbers.  A phonon in mode $k$ is created by the operator
$a_k^+$. To first order in phonon creation/destruction
operators the electron-phonon interaction is given by
$\gamma^k_{i,j}$, and this governs the process of an electron
hopping from site $i$ to site $j$ and exciting or de-exciting a
phonon in mode $k$.  The Hamiltonian of the $N$-site wire that
we consider is,
\begin{eqnarray*}
H_C &=& \sum_{i=1}^N e_i b_{i}^+ b_i + \sum_{i \ne j}
 t_{i,j}^{0} (b_{i}^+ b_j + \mathrm{h.c.}) +  \sum_k
 \sum_{i \ne j} \gamma^{k}_{i,j} (a_{k}^+ +
 a_k)(b_{i}^+ b_j + \mathrm{h.c.}) +  \sum_{k} \hbar
 \omega_k a_{k}^+ a_k
\end{eqnarray*}
The leads are modelled as a chain of atoms each with a single
atomic orbital $|n\rangle$, where $n$ labels the site.  For the
left lead $n$ ranges from $-\infty$ to $-1$.  On the right lead
$n$ ranges from $1$ to $\infty$.  On the left lead an electron
is created on site $n$ by the operator $c_n^+$. The site energy
is given by $\epsilon_L = \epsilon + eV/2$. Electrons can hop to
nearest neighbor sites and this is controlled by the hopping
parameter $\beta_L$.  Similarly on the right lead an electron
can exist on site $n$ in an orbital with energy $\epsilon_R =
\epsilon - eV/2$.  The nearest neighbor hopping is controlled by
the parameter $\beta_R$. Since we assume that there
are no phonons on the leads the Hamiltonians for the left
and right leads are given by
\begin{eqnarray*}
H_L &=& \epsilon_L \sum_{n=-\infty}^{1} c_{n}^+ c_n +
\beta_L \sum_{n=-\infty}^{1} (c_{n}^+ c_{n-1} +
\mathrm{h.c}) \\
H_R &=& \epsilon_R \sum_{n=1}^{\infty} c_{n}^+ c_n +
\beta_R \sum_{n=1}^{\infty} (c_{n}^+ c_{n-1} +
\mathrm{h.c})
\end{eqnarray*}
The wire is coupled to the leads via $V_L$ and $V_R$. We assume
that only the sites on the leads directly adjacent to the wire
are coupled to the wire. Electrons can hop from these lead
sites to any site $j$ on the wire and this is governed by
$w_{-1,j}$ for the left lead and $w_{1,j}$ for the right lead.
We take the electron-phonon interaction between the leads and
the wire to first order.  This allows for electrons from the
leads to hop onto the wire and excite a phonon in mode $k$.
This electron-phonon interaction is given by $\Gamma^k_{-1,j}$
for the left lead and $\Gamma^k_{1,j}$ for the right lead.
Thus the coupling potentials are,
\begin{eqnarray*}
V_L &=& \sum_{j=1}^{N} w_{-1,j}^0(c_{-1}^+ b_j +
\mathrm{h.c.}) + \sum_k \sum_{j=1}^{N}
\Gamma^{k}_{-1,j}(a_{k}^+ + a_k)(c_{-1}^+ b_j +
\mathrm{h.c.})\\
V_R &=& \sum_{j=1}^{N} w_{1,j}(c_{1}^+ b_j +
\mathrm{h.c.}) + \sum_k \sum_{j=1}^{N}
\Gamma^{k}_{1,j}(a_{k}^+ + a_k)(c_{1}^+ b_j +
\mathrm{h.c.})
\end{eqnarray*}

We now determine the scattering wavefunction from
Schroedinger's equation, $H |\Psi^{\alpha}\rangle = E
|\Psi^{\alpha}\rangle$, where $\alpha$ labels the initial
phonon state on the conducting wire. The wavefunction describes
the many body system consisting of a single electron and the
phonons. As shown by Bon$\breve{c}$a and Trugman, the
scattering process can be represented graphically.\cite{Hau98}
The power of this representation is in that the many-body
problem can be viewed as a multi-channel single-electron
scattering problem.  The $\alpha'$ channel in either the left
or the right leads corresponds to an electron propagating in
the single electron mode of that lead with the wire being in
its $\alpha'$ phonon state.  A scattering process which changes
the phonon state from $|\alpha\rangle = |\{n_k\}\rangle$ to
$|\alpha'\rangle=|\{n'_k\}\rangle$ can then be represented
graphically as an electron incident from the $\alpha$ channel
and then scattering into the $\alpha'$ channel. Fig.~\ref{fig2}
shows a graphical representation of the scattering channels for
a simple single site wire described by the above
Hamiltonian. An electron incident in the $\alpha$ channel in
the left lead can scatter elastically or inelastically into any
of the outgoing channels in the left and right lead. Channels
are connected horizontally by the electron hopping part of the
Hamiltonian whereas they are connected vertically by the
electron-phonon interaction.

With this picture in mind, we now write explicit forms for the
wavefunction on the left lead (L), right lead (R) and the wire
conductor (C). We assume that the wire is initially in the
phonon state $|\alpha\rangle$ and that there is a rightward
propagating electron in a Bloch mode $\sum_n e^{i n y^{\alpha}}
|n\rangle$ with energy $E$ in the left lead. The electron can
then be reflected elastically or inelastically into any of the
other channels in the left lead and propagate in a new Bloch
mode with energy $E'$ in the left lead with the wire in a new
phonon state $|\alpha'\rangle$. Thus, the wavefunction on the
left lead will consist of the initial state plus a sum over all
possible reflected states with some unknown coefficients
$r_{\alpha',\alpha}$. On the wire the wavefunction will be a
linear combination of the product of the atomic orbitals with
all of the different phonon states.  On the right lead the
electron can be transmitted into any of the outgoing channels
and propagate in a Bloch state with an energy $E'$, leaving the
wire in a phonon state $|\alpha'\rangle$. Thus the wavefunction
on the right lead will be a sum over all transmitted states
with coefficients $t_{\alpha',\alpha}$ to be determined. Thus
we have
\begin{eqnarray}
|\Psi^{\alpha}\rangle_L &=& \sum_{n=-\infty}^{-1}
e^{iny^{\alpha}} |n\rangle \otimes |\alpha\rangle +
\sum_{\alpha'} r_{\alpha',\alpha} \sum_{n=-\infty}^{-1}
e^{-iny^{\alpha'}} |n\rangle \otimes |\alpha'\rangle \\
|\Psi^{\alpha}\rangle_C &=& \sum_{\alpha'} \sum_{j=1}^N
u_j^{\alpha',\alpha} |j\rangle \otimes |\alpha'\rangle \\
|\Psi^{\alpha}\rangle_R &=&
\sum_{\alpha'} t_{\alpha',\alpha} \sum_{n=1}^{\infty}
e^{iny^{\alpha'}} |n\rangle \otimes |\alpha'\rangle
\end{eqnarray}
with
\begin{equation}
|\Psi^\alpha \rangle  = |\Psi^{\alpha}\rangle_L +
|\Psi^{\alpha}\rangle_C + |\Psi^{\alpha}\rangle_R
\nonumber
\end{equation}
The total energy of the system is conserved, so $E = E' +
\sum_k (n'_k - n_k) \hbar \omega_k$, where $E'$ is the energy
of the scattered electron. $y^{\alpha'}$ corresponds to the
reduced wavevector of an electron propagating with energy $E'$
and is determined from the condition $E' = \alpha_{L/R} +
2\beta_{L/R} \cos(y^{\alpha'})$ depending on whether it is in
the left or right lead.

Inserting the above wavefunction into Schr\"{o}dinger's equation
$H |\Psi^{\alpha}\rangle = E |\Psi^{\alpha}\rangle$ yields a
system of linear equations which can be solved numerically for
the $t_{\alpha',\alpha}$, $r_{\alpha',\alpha}$, and the
$u_j^{\alpha',\alpha}$ at each energy $E$. For our particular
model based on single channel leads we then determine the
transmission and reflection coefficients for each inelastic
channel using,
\begin{eqnarray}
T^{\alpha,\alpha'}_{L\rightarrow R}(E,E') &=& \left |
\frac{ v_R^{\alpha'} }{ v_L^{\alpha} }
\right | |t_{\alpha',\alpha}|^2 \\
R^{\alpha,\alpha'}_{L}(E,E') &=& \left | \frac{v_L^{\alpha'}}{v_L^{\alpha}}
\right | |r_{\alpha',\alpha}|^2
\end{eqnarray}
where $v_L^{\alpha'}$ and $v_R^{\alpha'}$ are the velocity of
the electron in the $\alpha'$ channel of the left or right lead
respectively.

We then solve the corresponding problem for electrons incident
from the right lead to determine
$T^{\alpha,\alpha'}_{R\rightarrow L}(E,E')$ and
$R^{\alpha,\alpha'}_{R}(E,E')$. We now proceed to show how
these quantities are used to calculate the non-equilibrium
electron distribution and current for the above system.

\section{Non-equilibrium electron distributions and the
evaluation of the electronic current}
As mentioned in the introduction, when calculating the current
using Landauer theory the Pauli exclusion principle for final
states does not play a role; scattered electrons do not compete
with each other.  A complication that arises when inelastic
scattering processes are considered is that electrons can now
compete with other electrons that are scattered to the same
final state; the Pauli principle becomes important. For
instance, for the system discussed in Sec. II, an electron
which suffers an inelastic collision that puts it into a final
state with energy $E'$ in the drain competes with all other
scattered electrons that can occupy that final state (as shown
in Fig.~\ref{fig3}). The likelihood that this state is
unoccupied in the drain is $(1 - f(E'))$, where $f(E')$ is the
non-equilibrium electron distribution function in the drain.
Using the equilibrium electron distribution (Fermi function)
here is not correct since we are discussing transport, an
inherently non-equilibrium phenomenon. We will now explain how
to determine this non-equilibrium distribution for the model
presented in Sec.~II.

Fig.~\ref{fig3} depicts the processes we are considering for
electrons that scatter from various initial states with
differing energies $E$ to the same final state at energy $E'$
in the right lead.  Some assumptions must be made about which
processes can compete with each other. To be consistent with
Landauer theory we assume that elastic scattering processes do
not compete with other elastic scattering processes (these
correspond to those processes which both begin and end at energy
$E'$ labelled A1 and B1 in Fig.~\ref{fig3}). With this
assumption an electron which is elastically scattered from the
source to the drain at energy $E'$ does not compete with an
electron which is elastically reflected back into the drain at
the same energy (process B1). However, elastically scattered
electrons can compete with all electrons that can be
inelastically scattered to the same final state. The
inelastically scattered electrons correspond to those electrons
which start at various energies $E$ labelled A2, A3 and B2 in
Fig.~\ref{fig3}. We also assume that inelastically scattered
electrons compete with all other scattered electrons that start
in a different initial electron state (i.e. electrons
undergoing process A2 compete with electrons suffering
processes A1, A3,  B1 and B2).

At a given final scattered energy $E'$, we consider the
following contributions to the non-equilibrium rightward moving
electron distribution in the right lead: elastically scattered
electrons transmitted from the left lead
$(f^{\alpha,\alpha}_{+,el})^R(E')$, inelastically scattered
electrons transmitted from the left lead
$(f^{\alpha,\alpha'}_{+,in})^R(E')$ (the final phonon state is
$\alpha'$), elastically scattered electrons which originate in
the right lead and are reflected back into the right lead
$(f^{\alpha,\alpha}_{-,el})^R(E')$, and inelastically scattered
electrons which start in the right lead and are reflected back
into the right lead $(f^{\alpha,\alpha'}_{-,in})^R(E')$. (The
$(+)$ denotes transmitted electrons, the $(-)$ reflected
electrons and the $R$ signifies that these are the
distributions in the right lead).

To determine the contribution to the non-equilibrium
distribution in the right lead {\em that is due to transmitted
electrons} we must first consider the distribution of
electrons incident from the left lead. As in Landauer theory, we
assume that this distribution is given by the equilibrium
electron distribution function which exists deep in the left
lead. This is the Fermi function for the left lead $F_L(E) =
1/(exp((E-\mu_L)/kT)+1)$ where the electro-chemical potential is
$\mu_L = \epsilon_F + eV/2$ with $\epsilon_F$ being the common
Fermi energy of the two leads and $V$ the applied bias
voltage. An electron incident with energy $E$ from the left
lead can then scatter elastically or inelastically into the
final state with energy $E'$ with a transmission probability
$T^{\alpha,\alpha'}_{L\rightarrow R}(E,E')$. (Elastic
scattering corresponds to $\alpha = \alpha'$ and $E =
E'$). Because we now have a number of processes which can place
electrons into this same final state the Pauli exclusion
principle must be considered. To incorporate this into our
determination of the distribution function we must then
consider the probability that this state is unoccupied by
competing electrons in the drain. For the distribution due to
elastically scattered electrons, this probability will be
determined by the non-equilibrium electron distributions of the
inelastically transmitted and reflected electrons, according to
the above assumptions. It is given by $(1 -
\sum_{\alpha'} (f^{\alpha,\alpha'}_{+,in})^R(E') -
\sum_{\alpha'} (f^{\alpha,\alpha'}_{-,in})^R(E') )$. For the
distribution of a specific inelastic transmitting process, the
probability that the state in the right lead is unoccupied by
competing electrons is determined by the distributions due to
all other scattering processes.  It is given by $(1 -
(f^{\alpha,\alpha}_{+,el})^R(E') -
(f^{\alpha,\alpha}_{-,el})^R(E') - \sum_{\alpha'}
(f^{\alpha,\alpha'}_{-,in})^R(E') - \sum_{\beta\ne \alpha'}
(f^{\alpha,\beta}_{+,in})^R(E') )$. The transmitted electron
distributions will be proportional to the product of the above
probabilities with $F_L$ and with the respective
transmission probability $T$.  We define the appropriate
proportionality constant for electrons incident from the left
lead with energy
$E$, to be $c(E)$. This
proportionality constant will be determined by imposing the
constraint of particle conservation.  Combining the above
considerations we arrive at the following expressions for the
contributions to the non-equilibrium electron distribution due
to elastically and inelastically transmitted electrons:
\begin{equation}
(f^{\alpha,\alpha}_{+,el})^R(E') = c(E') F_L(E')
T^{\alpha,\alpha}_{L\rightarrow R}(E',E') (1 - \sum_{\alpha'}
(f^{\alpha,\alpha'}_{+,in})^R(E') - \sum_{\alpha'}
(f^{\alpha,\alpha'}_{-,in})^R(E') )
\label{eqn:8}
\end{equation}
\begin{eqnarray}
(f^{\alpha,\alpha'}_{+,in})^R(E') = c(E) F_L(E)
T^{\alpha,\alpha'}_{L \rightarrow R}(E,E') (1 -
(f^{\alpha,\alpha}_{+,el})^R(E') -
(f^{\alpha,\alpha}_{-,el})^R(E') - \nonumber \\ \sum_{\alpha'}
(f^{\alpha,\alpha'}_{-,in})^R(E') - \sum_{\beta\ne \alpha'}
(f^{\alpha,\beta}_{+,in})^R(E') )
\end{eqnarray}
Similarly for the contributions to the non-equilibrium electron
distribution in the right lead due to elastically and
inelastically {\em reflected} electrons, we must consider the
distribution of incident electrons given by the Fermi function
$F_R(E)= 1/(exp((E-\mu_R)/kT) +1)$ for the right lead where
$\mu_R = \epsilon_F - eV/2$. Those electrons which are incident
with energy $E$ can be inelastically or elastically
reflected into the final state with energy $E'$ with a
probability
$R^{\alpha,\alpha'}_{R}(E,E')$. Again we take into account the
competition for the occupation of this state using the Pauli
principle. For elastically reflected electrons the probability
that this state is unoccupied is the same as that for
elastically transmitted electrons.  For inelastically reflected
electrons the probability that the state is unoccupied is
determined by all distributions due to different scattering
processes.  It is given by $(1 -
(f^{\alpha,\alpha}_{+,el})^R(E') -
(f^{\alpha,\alpha}_{-,el})^R(E') -
\sum_{\alpha'}(f^{\alpha,\alpha'}_{+,in})^R(E') - \sum_{\beta
\ne \alpha'} (f^{\alpha,\beta}_{-,in})^R(E') )$. Again, the
reflected electron distribution will be proportional to the
product of the above probabilities with R and $F_R$.
For electrons
incident from the right lead with energy $E$, the
proportionality constant is
$d(E)$.  Based on these considerations we write the
non-equilibrium distributions in the right lead due to
elastically and inelastically scattered electrons as
\begin{equation}
(f^{\alpha,\alpha}_{-,el})^R(E') = d(E') F_R(E')
R^{\alpha,\alpha}_{R}(E',E') (1 - \sum_{\alpha'}
(f^{\alpha,\alpha'}_{+,in})^R(E') - \sum_{\alpha'}
(f^{\alpha,\alpha'}_{-,in})^R(E') )
\end{equation}
\begin{eqnarray}
(f^{\alpha,\alpha'}_{-,in})^R(E') = d(E) F_R(E)
R^{\alpha,\alpha'}_{R}(E,E') (1 -
(f^{\alpha,\alpha}_{+,el})^R(E') -
(f^{\alpha,\alpha}_{-,el})^R(E') - \nonumber \\
\sum_{\alpha'}(f^{\alpha,\alpha'}_{+,in})^R(E') - \sum_{\beta
\ne \alpha'} (f^{\alpha,\beta}_{-,in})^R(E') )
\label{eqn:11}
\end{eqnarray}

Similarly a system of equations is set up for the
non-equilibrium electron distributions in the left lead,
$(f^{\alpha,\alpha}_{+,el})^L$,
$(f^{\alpha,\alpha'}_{+,in})^L$,
$(f^{\alpha,\alpha}_{-,el})^L$, and
$(f^{\alpha,\alpha'}_{-,in})^L$. (Here, $(+)$ signifies
transmitted electrons from the right lead to left lead, $(-)$
refers to reflected electrons in the left lead, and $(L)$
denotes that all these distributions are in the left lead).

The last consideration we must impose is particle
conservation. Particle conservation requires that the total
incoming current be equal to the sum of the total transmitted
current plus the total reflected current. This will allow us to
determine the constants of proportionality for both the left
and right leads. For the left lead we apply the following
equality,
\begin{equation}
F_L(E) = (f^{\alpha,\alpha}_{+,el})^R(E) +
(f^{\alpha,\alpha}_{-,el})^L(E) + \sum_{\alpha'}
(f^{\alpha,\alpha'}_{+,in})^R(E') + \sum_{\alpha'}
(f^{\alpha,\alpha'}_{-,in})^L(E') \label{eqn:equality}
\end{equation}
This equality expresses that the incident electron distribution
is equal to the sum over all transmitted and reflected
electron distributions that arise from that incident channel.
Since all of the transmitted and reflected distributions
originated from the same incident channel, they all involve the
same proportionality constant $c(E)$.  Inserting the
expressions for the $f$'s into Equation (\ref{eqn:equality}) allows
us to solve for
$c(E)$.  This yields,
\begin{equation}
c(E) = \frac{1}{A_{el} + A_{in} + B_{el} + B_{in}}
\label{eqn:c}
\end{equation}
where
\begin{eqnarray*}
A_{el} = T^{\alpha,\alpha}_{L\rightarrow R}(E,E) (1 -
\sum_{\alpha'} (f^{\alpha,\alpha'}_{+,in})^R(E) -
\sum_{\alpha'} (f^{\alpha,\alpha'}_{-,in})^R(E) ) \\
A_{in} =
\sum_{\alpha'} T^{\alpha,\alpha'}_{L \rightarrow R}(E,E') (1 -
(f^{\alpha,\alpha}_{+,el})^R(E') -
(f^{\alpha,\alpha}_{-,el})^R(E') - \\ \sum_{\beta}
(f^{\alpha,\beta}_{-,in})^R(E') - \sum_{\beta'\ne \alpha'}
(f^{\alpha,\beta'}_{+,in})^R(E') )\\
B_{el} = R^{\alpha,\alpha}_{L}(E,E) (1 - \sum_{\alpha'}
(f^{\alpha,\alpha'}_{+,in})^L(E) - \sum_{\alpha'}
(f^{\alpha,\alpha'}_{-,in})^L(E) ) \\
B_{in} = \sum_{\alpha'} R^{\alpha,\alpha'}_{L}(E,E') (1 -
(f^{\alpha,\alpha}_{+,el})^L(E') -
(f^{\alpha,\alpha}_{-,el})^L(E') - \\
\sum_{\beta}(f^{\alpha,\beta}_{+,in})^L(E') - \sum_{\beta'
\ne \alpha'} (f^{\alpha,\beta'}_{-,in})^L(E') )
\end{eqnarray*}
Similarly, $d(E)$ can be determined by applying a similar
equality for the distributions originating from the right
lead.  This yields,
\begin{equation}
d(E) = \frac{1}{C_{el} + C_{in} + D_{el} + D_{in}}
\label{eqn:d}
\end{equation}
where
\begin{eqnarray*}
C_{el} = T^{\alpha,\alpha}_{R\rightarrow L}(E,E) (1 -
\sum_{\alpha'} (f^{\alpha,\alpha'}_{+,in})^L(E) -
\sum_{\alpha'} (f^{\alpha,\alpha'}_{-,in})^L(E) ) \\
C_{in} =
\sum_{\alpha'} T^{\alpha,\alpha'}_{R \rightarrow L}(E,E') (1 -
(f^{\alpha,\alpha}_{+,el})^L(E') -
(f^{\alpha,\alpha}_{-,el})^L(E') - \\ \sum_{\beta}
(f^{\alpha,\beta}_{-,in})^L(E') - \sum_{\beta'\ne \alpha'}
(f^{\alpha,\beta'}_{+,in})^L(E') )\\
D_{el} = R^{\alpha,\alpha}_{R}(E,E) (1 - \sum_{\alpha'}
(f^{\alpha,\alpha'}_{+,in})^R(E) - \sum_{\alpha'}
(f^{\alpha,\alpha'}_{-,in})^R(E) ) \\
D_{in} = \sum_{\alpha'} R^{\alpha,\alpha'}_{R}(E,E') (1 -
(f^{\alpha,\alpha}_{+,el})^R(E') -
(f^{\alpha,\alpha}_{-,el})^R(E') - \\
\sum_{\beta}(f^{\alpha,\beta}_{+,in})^R(E') - \sum_{\beta'
\ne \alpha'} (f^{\alpha,\beta'}_{-,in})^R(E') )
\end{eqnarray*}

The above self-consistent system of non-linear equations must
be solved iteratively to determine the unknown distributions
$f$, and the unknown proportionality constants $c(E)$ and
$d(E)$. For an initial trial solution, we let the $f$'s
take on the following values,
\begin{eqnarray*}
(f^{\alpha,\alpha}_{+,el})^R(E) &=& \frac{F_L(E)
T^{\alpha,\alpha}_{L\rightarrow R}(E,E)}
{T^{\alpha,\alpha}_{L\rightarrow R}(E,E) +
R^{\alpha,\alpha}_{L}(E,E)} \\
(f^{\alpha,\alpha}_{-,el})^R(E) &=& \frac{F_R(E)
R^{\alpha,\alpha}_{R}(E,E)}{T^{\alpha,\alpha}_{R\rightarrow
L}(E,E) + R^{\alpha,\alpha}_{R}(E,E)} \\
(f^{\alpha,\alpha}_{+,el})^L(E) &=& \frac{F_R(E)
T^{\alpha,\alpha}_{R\rightarrow
L}(E,E)}{T^{\alpha,\alpha}_{R\rightarrow L}(E,E) +
R^{\alpha,\alpha}_{R}(E,E)} \\
(f^{\alpha,\alpha}_{-,el})^L(E)
&=& \frac{F_L(E)
R^{\alpha,\alpha}_{L}(E,E)}{T^{\alpha,\alpha}_{L\rightarrow
R}(E,E) + R^{\alpha,\alpha}_{L}(E,E)} \\
(f^{\alpha,\alpha'}_{\pm,in})^R(E) &=& 0 \\
(f^{\alpha,\alpha'}_{\pm,in})^L(E) &=& 0
\end{eqnarray*}
Using this initial guess for the $f$'s, we evaluate the
constants of proportionality $c(E)$ and $d(E)$ from Equations
\ref{eqn:c} and \ref{eqn:d}.  With these determined, we then
evaluate a new set of $f$'s from Equations
\ref{eqn:8}-\ref{eqn:11} using these
$c$'s and $d$'s and the old $f$'s.  We then iterate until the
$f$'s have converged.

With the non-equilibrium electron distributions thus determined
we proceed to evaluate the electronic current flowing through
the wire for an applied voltage $V$ and at a given temperature
$T$. At a given energy $E$ the current, $(\delta i_+)^R$,
flowing in an energy interval $\delta E$ due to rightward
propagating transmitted electrons in the right lead is given by
the charge $-e$ times the velocity $v_R(E)$, times the density
of states $dg(E)/dE$, times the electron distribution for
transmitted rightward propagating electrons in the right lead
$(f_{+,tot})^R(E) = (f^{\alpha,\alpha}_{+,el})^R(E) +
\sum_{\alpha'} (f^{\alpha,\alpha'}_{+,in})^R(E)$. Thus we can
write (including a factor of 2 for spin)
\begin{equation}
(\delta i_+)^R = -\frac{2 e}{h} (f_{+,tot})^R(E)\, \delta E
\label{eqn:dI}
\end{equation}
where in one dimension the density of states cancels the
velocity factor.  The total current due to rightward
propagating transmitted electrons in the right lead is found by
integrating Equation \ref{eqn:dI}, yielding,
\begin{equation}
(i_{+})^R(V) = - \frac{2e}{h} \int dE\;(f_{+,tot})^R(E)
\end{equation}
Similarly, the total current in the left lead due to
transmitted leftward propagating electrons is given by,
\begin{equation}
(i_{+})^L(V) = \frac{2e}{h} \int dE\;(f_{+,tot})^L(E)
\end{equation}
where $(f_{+,tot})^L(E) = (f^{\alpha,\alpha}_{+,el})^L(E) +
\sum_{\alpha'} (f^{\alpha,\alpha'}_{+,in})^L(E)$.
The net current flowing through the mesoscopic conductor
is then
\begin{equation}
i_{tot}(V) = (i_{+})^R(V) + (i_{+})^L(V)
\label{eq:itot}
\end{equation}

\section{Application to a diatomic molecule}
We now proceed to apply the above formalism to a simple
molecular wire system. It consists of two identical 1D leads
which are attached to the left and right ends of a two atom
chain.  Fig.~\ref{fig4} shows a schematic of our system. Each
atom making up the molecule has just one atomic orbital, so the
isolated two atom molecule has two discrete electronic energy
levels. When this molecule is coupled to the leads, incident
electrons with energies that roughly coincide with the energy
of either of these two states will be favourably
transmitted. In the absence of phonons we expect two strong
peaks in the transmission spectrum at the energies of the two
molecular levels. When longitudinal phonons on the molecule are
considered we expect interesting physics to occur when one of
the phonon energies corresponds to the energy gap between the
two energy levels of the molecule. When an electron is incident
at an energy corresponding to the higher molecular level it can
emit a phonon and drop to the lower electronic energy level.
This process should have a significant transmission probability
because the electron will be resonant with the higher energy
level and then also with the lower energy level. (Processes
where phonons are created and the electron is not resonant with
either of the two molecular energy levels will not be strongly
transmitting).  If we assume that the leads' Fermi energy lies
between the two molecular energy levels, we expect the
following behavior for the non-equilibrium electron
distributions: For low voltages and temperatures, the
non-equilibrium electron distribution in the drain will be
dominated by resonant electrons elastically scattered through
the lower energy level and elastically reflected electrons. As
the voltage increases the electro-chemical potential in the
left lead rises while it decreases in the right lead.
Eventually, at sufficiently high voltages there will be
electrons in the left lead with enough energy to undergo the
inelastic scattering process mentioned above.  The
non-equilibrium electron distribution in the drain will then
have a significant contribution due to these inelastically
scattered electrons. These inelastically scattered electrons
will compete with elastically scattered electrons that are
transmitted through the lower energy state. This competition
between electrons due to the Pauli principle will play an
important role in determining the current of this two site
system at higher voltages.

We have chosen the following set of
tight-binding parameters for the system described above. On the
leads, the site energy is $\epsilon = -10$ eV and the nearest
neighbor hopping energy is $\beta_L=\beta_R = -2$ eV.  For the
molecule, the two atoms have the same site energy $e_1 = e_2 =
-10.0$ eV and the hopping energy is $t_{1,2} = -0.15$ eV. The
coupling of the left lead to the first molecular site is
$w_{-1,1}=-0.2$ eV and the coupling of the second molecular
site to the right lead is $w_{2,1} = -0.2$ eV. We use the SSH
Hamiltonian to model the vibrational modes of the molecule and
electron-phonon coupling as is discussed in the Appendix: Our
vibrational model of the molecule has two longitudinal phonon
modes with energies
$\hbar \omega_1 = 0.24$ eV and $\hbar \omega_2 = 0.33$ eV.  The
electron-phonon (e-p) couplings on the chain are
$\gamma^1_{1,2} = 0.0$ eV and $\gamma^2_{1,2} = -0.24$ eV. The
e-p couplings between the leads and the chain are
$\Gamma^{1}_{-1,1} = 0.12$ eV, $\Gamma^{2}_{-1,1} = 0.12$ eV,
$\Gamma^1_{2,1} = -0.12$ eV and $\Gamma^2_{2,1} = 0.12$ eV. We
assume that the system is at a temperature of $T = 77$ K.

For the transport calculation, the initial phonon state
$|\alpha\rangle = |\{n^k\}\rangle$ is taken to be the average
phonon distribution on the chain for temperature $T$, with $n^k
= 1/(exp(\hbar \omega_k/(k_B T) ) - 1)$. We approximate each
average mode occupation number $n^k$ by the nearest integer.
At the temperature of interest, $|\alpha\rangle = |\{n_1,
n_2\}\rangle = |\{0,0\}\rangle$. The other phonon states that
we include in the calculation are those obtained by taking all
$n_i = 0,\ldots,N_{max}$ where $N_max = 2$.

In Fig.~\ref{fig5}a, the total transmission
probability found by summing over all outgoing channels (solid
line) is shown along with the transmission probability in the
absence of phonons (dashed line).  (These transmission
probabilities were calculated for 0 $V$ applied across the
molecule. We emphasize that they are the {\em one-electron}
transmission probabilities defined in Section II and do not
reflect the effects of the Pauli principle discussed in
Section III. The latter is taken into account in the
calculated non-equilibrium distributions and currents presented
later in this Section).  In the absence of phonons the two
strong resonant tunnelling peaks are seen. With the inclusion of
phonons, there are extra peaks due to the creation of phonons.
The two resonant peaks are shifted because the inclusion of
vibrations alters the hopping between sites which in turn
shifts the energies at which electrons are resonant. In
Fig.~\ref{fig5}b the transmission probability for elastically
scattered electrons is shown. The elastic spectrum show two
strong peaks at the molecular energy levels.  It also has other
smaller peaks which are due to the interaction of the electrons
with virtual phonons.

Fig.~\ref{fig6} shows the inelastic one-electron transmission
probability for three of the inelastic channels vs. the energy
$E$ of the incident electron state. The first graph,
Fig.~\ref{fig6}a is for the process where a phonon with $\hbar
\omega_1 = 0.24$ eV is created on the molecule.  The peak
labelled A corresponds to an electron emitting this phonon and
dropping into the lower resonant molecular energy level at
-10.26 eV.  The peak labelled B corresponds to resonant
tunnelling through the higher molecular energy level after the
creation of this phonon.  The two tunnelling peaks C and D at
the energies of the molecular resonant states correspond to an
electron entering the resonant molecular energy levels prior to
emitting the same phonon. The second graph, Fig.~\ref{fig6}b
corresponds to the creation of a single excitation of the
second phonon mode with energy $\hbar \omega_2 = 0.33$ eV,
equal to the molecular energy level gap. As discussed
previously, we expect electrons with energies incident on the
higher energy molecular state to be strongly transmitting and
this process corresponds to the peak labelled A in this graph.
Graph (c) is for the process where a single excitation is
created in both phonon modes.  The transmission probability of
this process is essentially the superposition of
Fig.~\ref{fig6}a and b, with an overall reduction in amplitude
because this corresponds to a higher order process.

We now proceed to calculate the non-equilibrium electron
distributions from the calculated one-electron transmission and
reflection coefficients using the method
presented in Sec. III.  We take the Fermi energy of the leads
to be $\epsilon_F = -10$ eV.  The non-equilibrium distributions
play an important role in the calculation of the current
because of the Pauli principle.  For instance, the presence of
inelastically transmitted or back-scattered electrons can
result in a significant suppression of the elastically
transmitted current.  In this system, as mentioned above, the
process whereby an electron creates a phonon with energy $\hbar
\omega_2$ and is then at resonance with the lower electronic
resonant tunnelling state, is strongly transmitting in the
inelastic spectrum at
$E=-9.9$eV.  At low voltages there are not very many of
these particular inelastic processes occurring since there are
not many electrons with this energy incident from the left
lead, and the final electron states for the transitions are
almost fully occupied by competing elastic processes. At higher
voltages (greater than 0.55 V) there is a significant number of
these processes and the non-equilibrium distribution
$f^{\alpha,\alpha'}_{+,in}(E)$ where $\alpha'$ corresponds to
the phonon state $\{n'_k\} =
\{0,1\}$ is significant.  As the voltage increases further, the
total distribution due to inelastically scattered electrons
continues to grow as more processes become energetically
favourable. The total non-equilibrium Fermi distribution for
all of the transmitted elastically scattered electrons is
shown in Fig.~\ref{fig7}a for 0.5 V (solid line) and 1 V (dashed
line). Fig.~\ref{fig7}b shows the corresponding distribution
for inelastically transmitted electrons. It highlights how this
distribution function increases as more inelastic processes
become energetically favourable. It also shows that there is a
significant presence of inelastically scattered electrons in
the drain for energies around -10.25 eV. At this energy,
elastically scattered electrons are also particularly strongly
transmitting and the dip in the distribution of the
inelastically scattered electrons at this energy is due to the
presence of these elastically scattered electrons in the drain.
This is labelled ``A'' in Fig.~\ref{fig7}b.

These non-equilibrium distributions are now used to evaluate
the current flowing through the two atom molecule.  In
Fig.~\ref{fig8}a,b we show the total rightward flowing current
that is elastically and inelastically transmitted through the
molecule, evaluated using our method.

We also show (dotted lines) the corresponding currents
evaluated assuming instead that the non-equilibrium
distribution of electrons in the drain that appears in the
Pauli exclusion factor is replaced simply by the equilibrium
Fermi function in the right lead.  For transmitted elastically
scattered electrons this simplified expression for the current
is
\begin{equation}
i^+_{el} = - \frac{2 e}{h} \int dE \;
T^{\alpha,\alpha}_{L\rightarrow R}(E,E) F_L(E) (1 - F_R(E)) \label{eqn:eqlb}
\end{equation}
For models that do {\em not} include electron-phonon
interactions, the {\em net} current through the molecule
obtained in this way is the same\cite{Hau98} as is obtained
from Landauer theory, although the factor $(1-F_R(E))$ does not
appear in the latter as was explained in the
introduction. However if electron-phonon interactions are
present, approximating the effects of the Pauli principle in
this way does not account for competition between elastic and
inelastic processes which (at higher voltages) inject electrons
into states that are unoccupied according to the equilibrium
distribution. This approximation (even at low voltages) also
violates particle conservation when electron-phonon
interactions are present.

By using the non-equilibrium distributions described in Section
III these effects are properly taken into account and current
is conserved. All of this is reflected in the graphs: In
Fig.~\ref{fig8}a initially the solid curve (the present result)
shows significantly more elastic current (as expected from
Landauer theory), and then eventually the elastic current is
reduced due to the presence of transmitted inelastically
scattered electrons in the normally unoccupied states of the
drain. The total current flowing through the molecule as a
function of voltage is shown in Fig.~\ref{fig8}c. The larger
total current exhibited by the solid curve is a manifestation
of current conservation in the present formalism.  The dashed
curve falls below the solid curve because the approximations
used to calculate it do not conserve current.

\section{Conclusions}
We have shown that to generalize Landauer theory to include the
effects of inelastic scattering, it is necessary to calculate
self-consistently the non-equilibrium contributions to the
electron distribution in the source and the drain leads.  We
have developed a method for determining these non-equilibrium
distributions and applied it to a simple model.  The model was
based on single-channel ideal leads attached to a mesoscopic
conductor.  Phonons were only considered on the conductor. The
non-equilibrium distributions were determined solely in terms
of the Fermi functions of the left and right leads and the
transmission and reflection probabilities for the various
one-electron elastic and inelastic scattering processes
considered. We showed how to properly take account of the Pauli
exclusion principle and particle conservation in the
calculation of the non-equilibrium electron distribution. We
applied our approach to a two site molecule. One of the phonons
had an energy that corresponded to the molecular energy level
gap. For this system we found that the details of the
non-equilibrium electron distribution had an important role in
the determination of the current.  At higher voltages we found
that the distribution in the right lead due to inelastically
scattered electrons began to compete significantly with
elastically scattered electrons because of the Pauli exclusion
principle. We found that this competition resulted in a
reduction of the elastically scattered current at higher
voltages.

\section{Acknowledgement}
This work was supported by NSERC.

\section{Appendix}
To find the vibrational modes of a molecule the atoms and bonds
may be modelled by a set of balls and springs. For our model,
we consider chain-like molecules, and represent the chain as a
set of balls with effective springs connecting nearest neighbor
atoms. The longitudinal displacement from equilibrium of the
$i^{th}$ atom with mass $m_i$ in the molecule is given by the
coordinate $q_i$. We assume that each atom only interacts with
its nearest neighbours via an effective spring coupling
$V_{i,j}$. The molecule has a discrete set of longitudinal
normal modes which vibrate with a particular frequency
$\omega_k$ and have atomic displacements given by $q^k_i =
d^k_i \exp{(-i \omega_k t)}$.  These normal modes satisfy the
familiar eigenvalue problem,
\begin{equation}
\sum_j V_{i,j} d_j^k = \omega_k^2 m_i d_i^k
\end{equation}
where for our model, the sum over $j$ is just over the nearest
neighbors of atom $i$.  Solving this equation yields the
eigen-frequencies and eigen-vectors for the longitudinal normal
modes.

The problem is canonically quantized by assuming $[q_i,p_j]=i
\hbar \delta_{i,j}$, where $p_j$ is the conjugate momentum to
$q_j$. We transform into normal mode operators $Q_k$ and $P_k$
by writing $q_i= \sum_k d^k_i Q_k$ and $p_i = \sum_k d^k_i m_i
P_k$. It is easily verified that $[Q_k,P_{k'}]= i\hbar
\delta_{k,k'}$. Performing second quantization on these
operators gives,
\begin{eqnarray}
Q_k = \sqrt{ \frac{\hbar}{2 \omega_k} }(a_k + a^+_k) \\
P_k = i \sqrt{ \frac{\hbar \omega_k}{2}}(a^+_k - a_k)
\end{eqnarray}
where $a^+_k$ and $a_k$ are creation and destruction operators
for the normal mode (or phonon) with frequency $\omega_k$
respectively.

The effects of longitudinal vibrations on the electronic
properties of the molecule can be taken into account by using
the Su-Schreiffer-Hieger (SSH) model.\cite{Su79} In it, the
nearest neighbor hopping parameter $t_{i,j}$ between two
neighboring atomic sites is expanded to linear order in terms
of their atomic displacements from equilibrium. This gives,
\begin{equation}
t_{i,j} = t_{i,j}^0 - \alpha_{i,j}(q_i - q_j)
\end{equation}
where $t_{i,j}^0$ is the hopping parameter at $q = q_i - q_j =
0$ and $\alpha_{i,j} = (d t_{i,j}/dq)_{q=0}$. (Note
$\alpha_{j,i} = -\alpha_{i,j}$).

By using the above expression for $q_i$ in terms of the
$Q_k$'s, and writing them in their second quantized form, this
gives $t_{i,j}$ as,
\begin{equation}
t_{i,j} = t_{i,j}^0 - \alpha_{i,j} \sum_k
(d_{i}^k-d_{j}^k)\sqrt{\frac{\hbar}{2w_k}}(a_k + a_{k}^+)
\end{equation}
This defines the mode dependent electron-phonon coupling,
$\gamma^{k}_{i,j}$ as,
\begin{equation}
\gamma^{k}_{i,j} = \alpha_{i,j}
\sqrt{\frac{\hbar}{2w_k}}(d_{i}^k-d_{j}^k)
\end{equation}
in terms of the known quantities $\alpha_{i,j}$, $\omega_k$ and
the $d^k$'s.

We now proceed to evaluate the eigen-frequencies and
eigen-vectors of the longitudinal modes for our two atom
molecule attached to two semi-infinite leads.  The atoms in the
leads are assumed fixed in their static positions.  The
effective spring coupling between the leads and the nearest
neighbor atoms on the molecule is given by $K = 180
eV/\AA^2$. The effective spring coupling between the two atoms
of the molecule is $k = 90 eV/\AA^2$. Each atom of the molecule
has a mass $m = 13$ A.M.U.. Solving the eigenvalue problem
using the above parameters we find two phonon modes with
energies of $\hbar \omega_1 = 0.24$ eV and $\hbar \omega_2 =
0.33$ eV.  The corresponding eigen-vectors are $d^1_1 = 0.054
\AA$, $d^1_2 = 0.054 \AA$ and $d^2_1 = 0.054 \AA$, $d^2_2 =
-0.054 \AA$.

In the SSH model the electrons are coupled to the phonons via
the parameter $\alpha_{i,j}$.  The nearest neighbor coupling
between the leads and the molecule and between the two sites on
the molecule is assumed to be $\alpha_{i,j} = 25$ eV/\AA for
$i$ to the right of $j$.  This gives the couplings between the
first site on the molecule and the left lead as
$\Gamma^1_{1,-1}=\Gamma^2_{1,-1} = 0.12$. The couplings between
the right lead and the second site on the molecule are
$\Gamma^1_{1,2} = -0.12$ and $\Gamma^2_{1,2} = 0.12$.  The
electron-phonon couplings on the molecule are $\gamma^1_{2,1} =
0$ and $\gamma^2_{2,1} = -0.24$.

\begin{figure}
\caption{A schematic of our model for a one dimensional
mesoscopic conductor.  The conductor bridges two one
dimensional ideal leads. The applied bias voltage is $V$. The
Hamiltonians of the individual elements making up the system
are labelled.}
\label{fig1}
\end{figure}
\begin{figure}
\caption{A graphical representation of the inelastic scattering
problem for a single site wire with first order electron-phonon
coupling between the leads and the wire .  Each phonon state of
the wire along with the Bloch state of the electron in a lead
can be visualized as a separate scattering channel.  Scattering
channels are connected horizontally by the electron hopping
parameters $t$ and $\beta$, while they are connected vertically
by the electron-phonon interaction $\Gamma, \gamma$.  An
electron is initially incident in the $\alpha$ channel and can
scatter into any of the other channels.}
\label{fig2}
\end{figure}
\begin{figure}
\caption{A schematic of the various elastic and inelastic
scattering processes that can occur within the mesoscopic
conductor that contribute to the rightward propagating electron
distribution in the right lead. The applied bias voltage is
$V$. The electrochemical potentials of the two leads under this
applied bias are labelled $\mu_L$ and $\mu_R$ for the left and
right leads respectively.  They share a common Fermi energy
labelled by $\epsilon_F$. Scattering processes originating in
the left lead are labelled A1-A3, and for those originating in
the right lead, they are labelled B1-B2.}
\label{fig3}
\end{figure}
\begin{figure}
\caption{A schematic of our two-site molecular wire.  We model
the longitudinal vibrations of the molecule using a nearest
neighbor ball and spring model.  The two atoms of the molecule
are coupled by an effective spring coupling $k$, while they are
coupled to the leads with a spring coupling $K$.}
\label{fig4}
\end{figure}
\begin{figure}
\caption{(a) Total transmission probability as a function of
energy $E$ of the incident electron for the two site
molecule. The dashed curve corresponds to the transmission
probability in the absence of phonons. The solid curve is with
the inclusion of longitudinal phonons in the calculation
at a temperature of $T=77$ K.
(b) The transmission probability for the elastic scattering
channel of the two site.  Both graphs are at $V = 0$.
molecule.}
\label{fig5}
\end{figure}
\begin{figure}
\caption{Transmission probabilities for three of the inelastic
scattering channels of the two site molecule. (a) Scattering
from the initial state $|\alpha\rangle = |0,0\rangle$ to
$|\alpha'\rangle = |1,0\rangle$. (b) Scattering from the initial
state $|\alpha\rangle = |0,0\rangle$ to $|\alpha'\rangle =
|0,1\rangle$. (c) Scattering from the initial
state $|\alpha\rangle = |0,0\rangle$ to $|\alpha'\rangle =
|1,1\rangle$. All graphs were calculated at $T = 77$ K and $V =
0$.}
\label{fig6}
\end{figure}
\begin{figure}
\caption{Non-equilibrium electron distributions in the right
lead for the two site molecule.  (a) Total non-equilibrium
electron distribution for transmitted elastically scattered
electrons at 0 $V$ (dashed curve) and 1 $V$ (solid curve). (b)
Total non-equilibrium electron distribution for transmitted
inelastically scattered electrons at 0.5 $V$ (solid curve) and
1 $V$ (dashed curve). The postion labelled ``A'' shows how
there is a dip in the inelastic distribution due to the sharp
rise in the elastic distribution at that energy $E$.  }
\label{fig7}
\end{figure}
\begin{figure}
\caption{(a) Total rightward flowing electronic current in the
right lead due to elastically transmitted electrons. (b) Total
rightward flowing electronic current in the right lead due to
inelastically transmitted electrons.  (c) The net electronic
current flowing through the two site molecule. The solid curves
are calculated from the non-equilibrium electron distribution
and the dashed curves are calculated using
Equation \ref{eqn:eqlb}. The current is in units of $2 e^2/h eV$.}
\label{fig8}
\end{figure}



\begin{references}
\bibitem{Stip1} B. C. Stipe, M. A. Rezaei and W. Ho,
Phys. Rev. Lett. {\bf 81}, 1263 (1998).
\bibitem{Stip2} B. C. Stipe, M. A. Rezaei and W. Ho, Science
{\bf 280}, 1732 (1998).
\bibitem{Stip3} B. C. Stipe, M. A. Rezaei, W. Ho, S. Gao,
M. Persson, and B. I. Lundqvist, Phys. Rev. Lett. {\bf 78},
4410 (1997).
\bibitem{Win88} N. S. Wingreen, K. W. Jacobsen, and
J. W. Wilkins, Phys. Rev. Lett. {\bf 61}, 1396 (1988).
\bibitem{Gelf89} B. Y. Gelfand, S. Schmitt-Rink and
A. F. J. Levi, Phys. Rev. Lett. {\bf 62}, 1683 (1989).
\bibitem{Stov91} J. A. St$\o$vneng, H. Hauge, P. Lipavsky, and
V. $\breve{S}$pi$\breve{c}$ka, Phys. Rev. B {\bf 44}, 13595
(1991).
\bibitem{Turl91} P. J. Turley and S. W. Teitsworth,
Phys. Rev. B. {\bf 44}, 3199 (1991).
\bibitem{Pers87} B. N. J. Persson and A. Baratoff,
Phys. Rev. Lett. {\bf 59}, 339 (1987).
\bibitem{Bin85} G. Binnig, N. Garcia, and H. Rohrer,
Phys. Rev. B {\bf 32}, 1336 (1985).
\bibitem{Gata93} M. A. Gata and P. R. Antoniewicz, Phys. Rev. B
{\bf 47}, 13797 (1993).
\bibitem{Hau98} K. Haule and J. Bon$\breve{c}$a, Phys. Rev. B
{\bf 59}, 13087 (1998).
\bibitem{Lan57} R. Landauer, IBM J. Res. Dev. {\bf 1}, 223
(1957); R. Landauer, Phys. Lett. {\bf 85A}, 91 (1981).
For recent reviews of Landauer theory see Y. Imry,
``Introduction to Mesoscopic Physics," Oxford University
Press, 1997, S. Datta, ``Electronic transport in
mesoscopic systems," Cambridge University Press, 1995.
\bibitem{Bonc95} J. Bon$\breve{c}$a and S. A. Trugman,
Phys. Rev. Lett. {\bf 75}, 2566 (1995).
\bibitem{Hol59} T. Holstein, Ann. Phys. (N.Y.) {\bf 8}, 325
(1959); {\bf 8}, 343 (1959).
\bibitem{Bonc97} J. Bon$\breve{c}$a and S. A. Trugman,
Phys. Rev. Lett. {\bf 79}, 4874 (1997).
\bibitem{Su79} W. P. Su, J. R. Schrieffer and A. J. Heeger,
Phys. Rev. Lett. {\bf 42}, 1698 (1979).
\bibitem{Ness99} H. Ness and A. J. Fisher,
Phys. Rev. Lett. {\bf 83}, 452 (1999).
\bibitem{Peierls1} See R. E. Peierls, ``Quantum Theory of
Solids," p.116, Oxford University Press (1964).
\bibitem{Peierls2} {\em Ibid.,} p.127.
\bibitem{Bal}For a general proof see L. E. Ballentine,
``Quantum Mechanics," p.332, Prentice-Hall (1990).
\end{references}
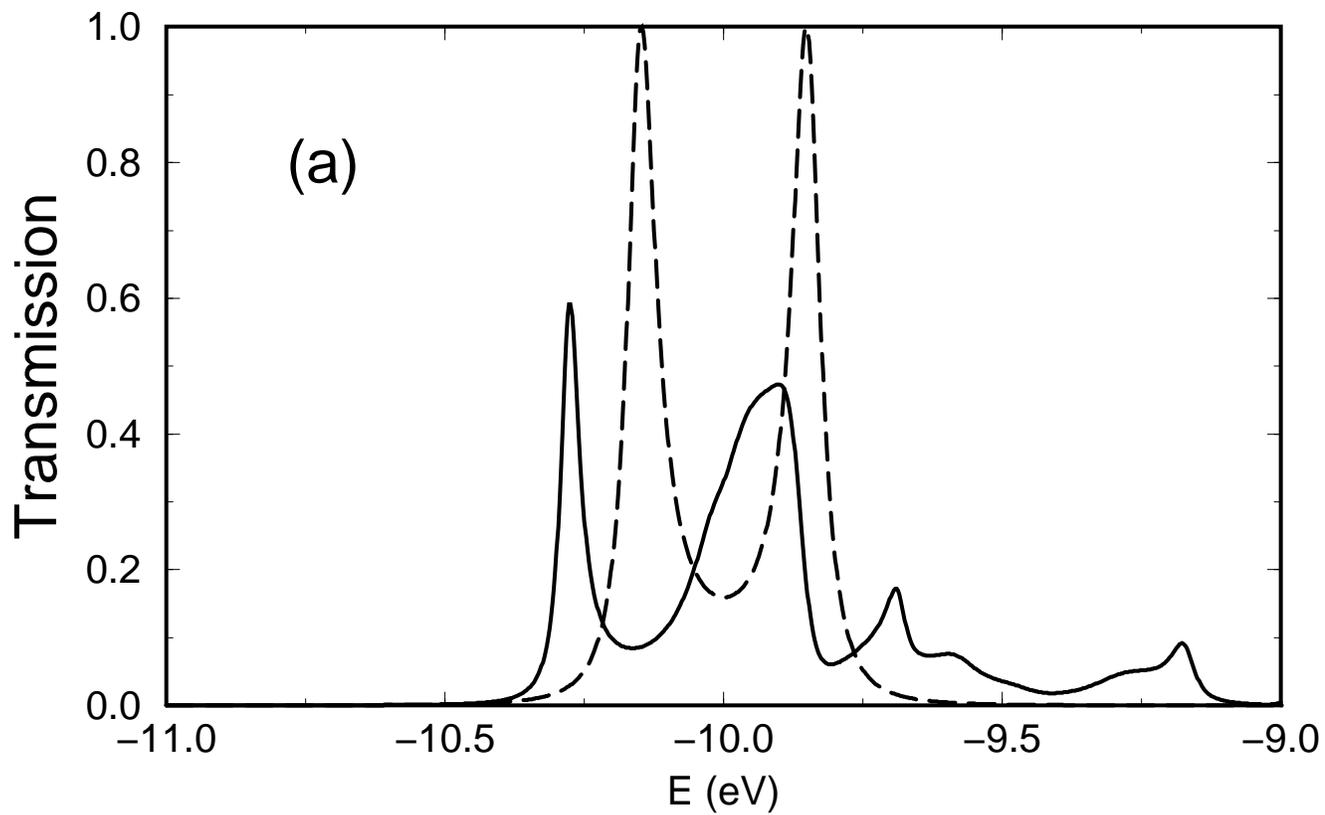
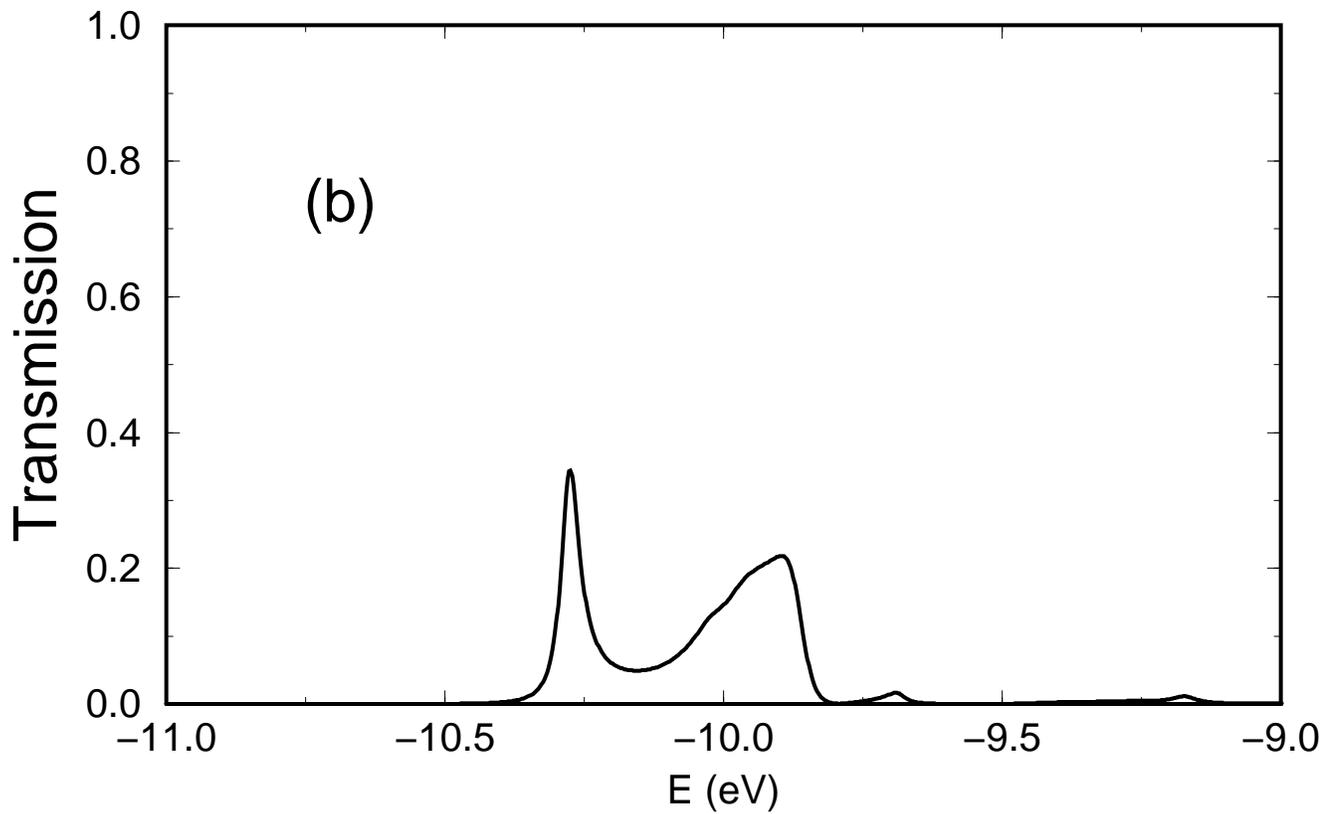
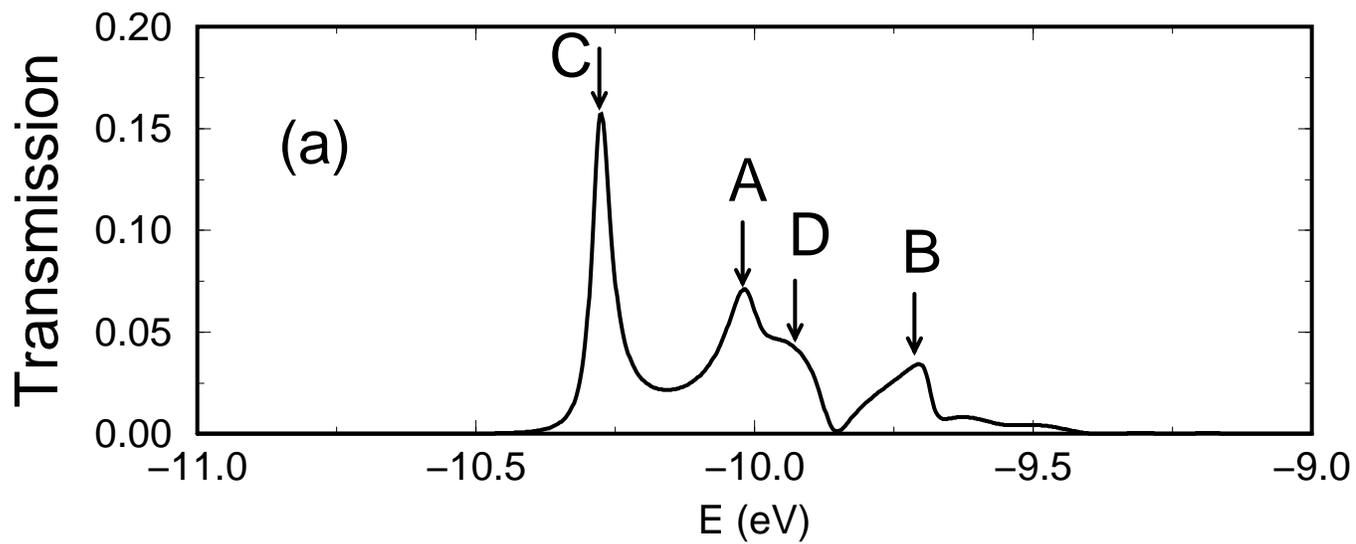
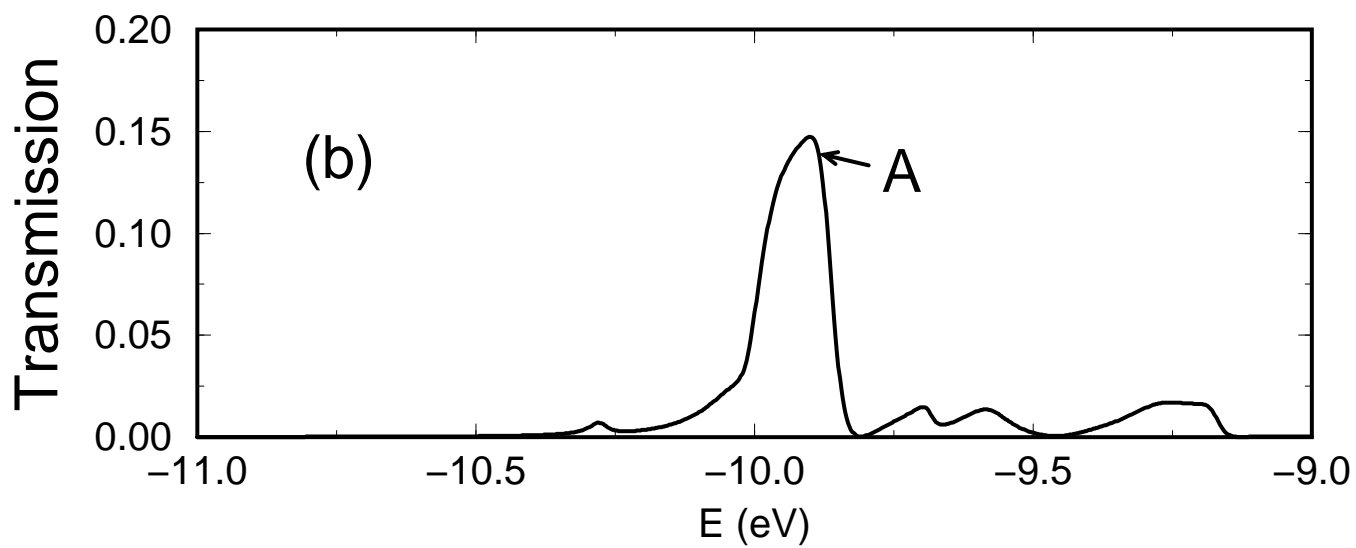
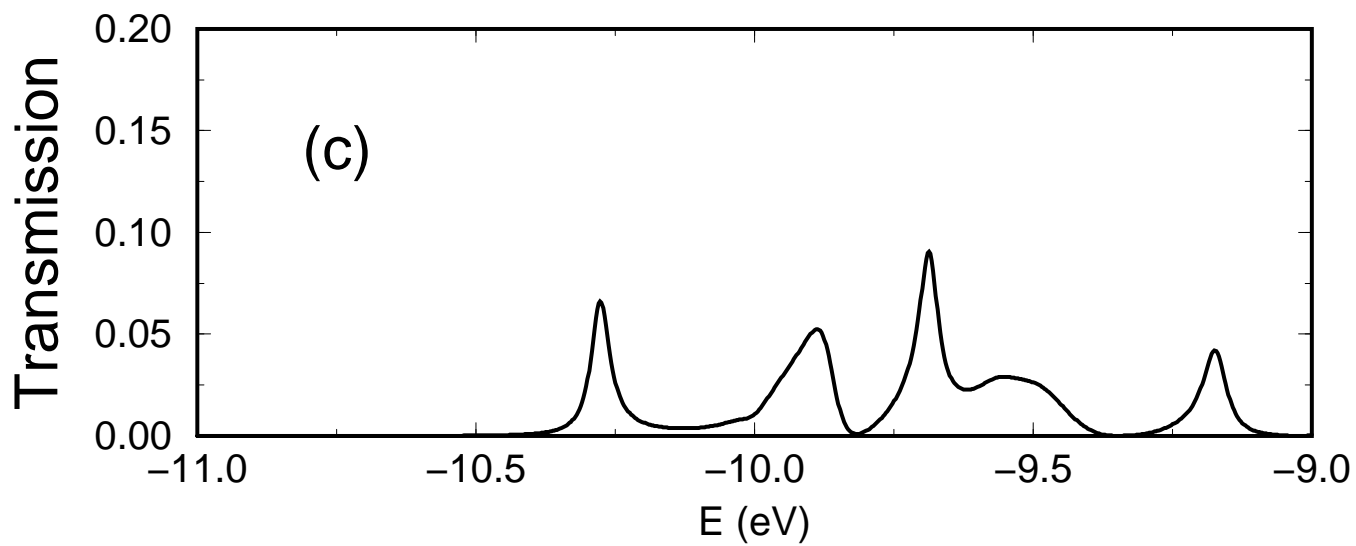
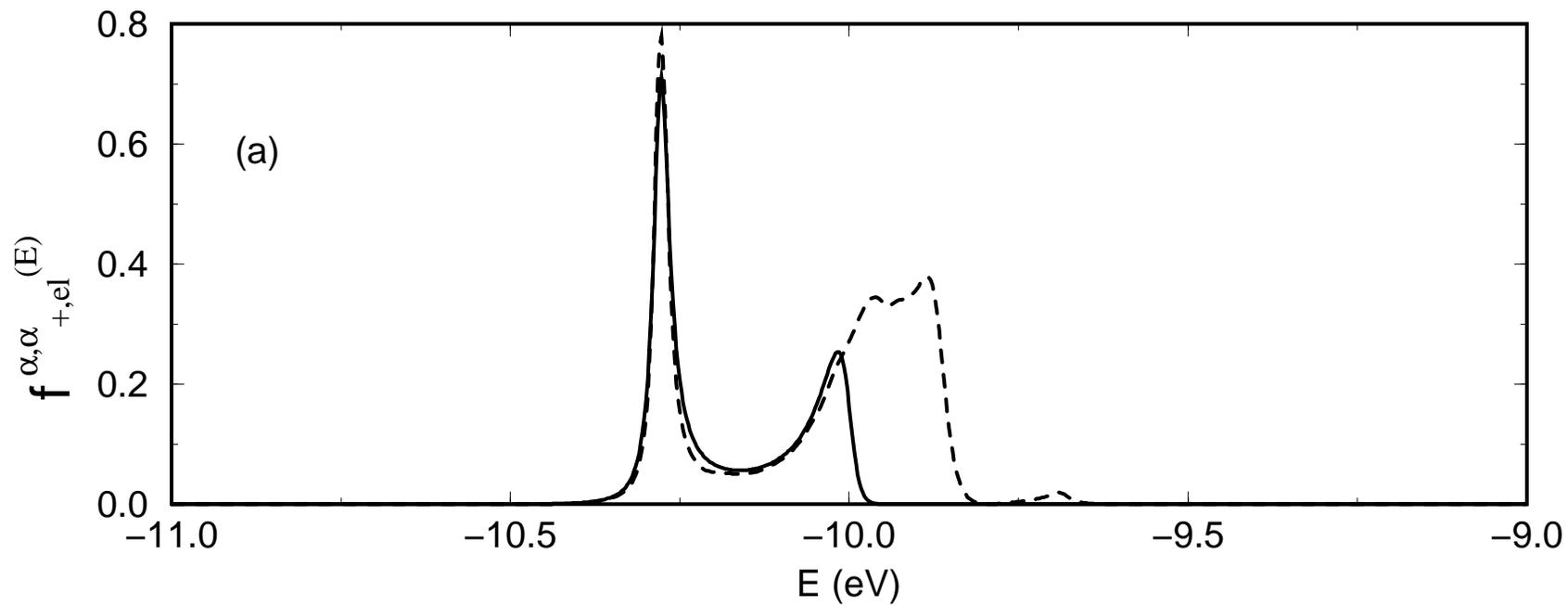
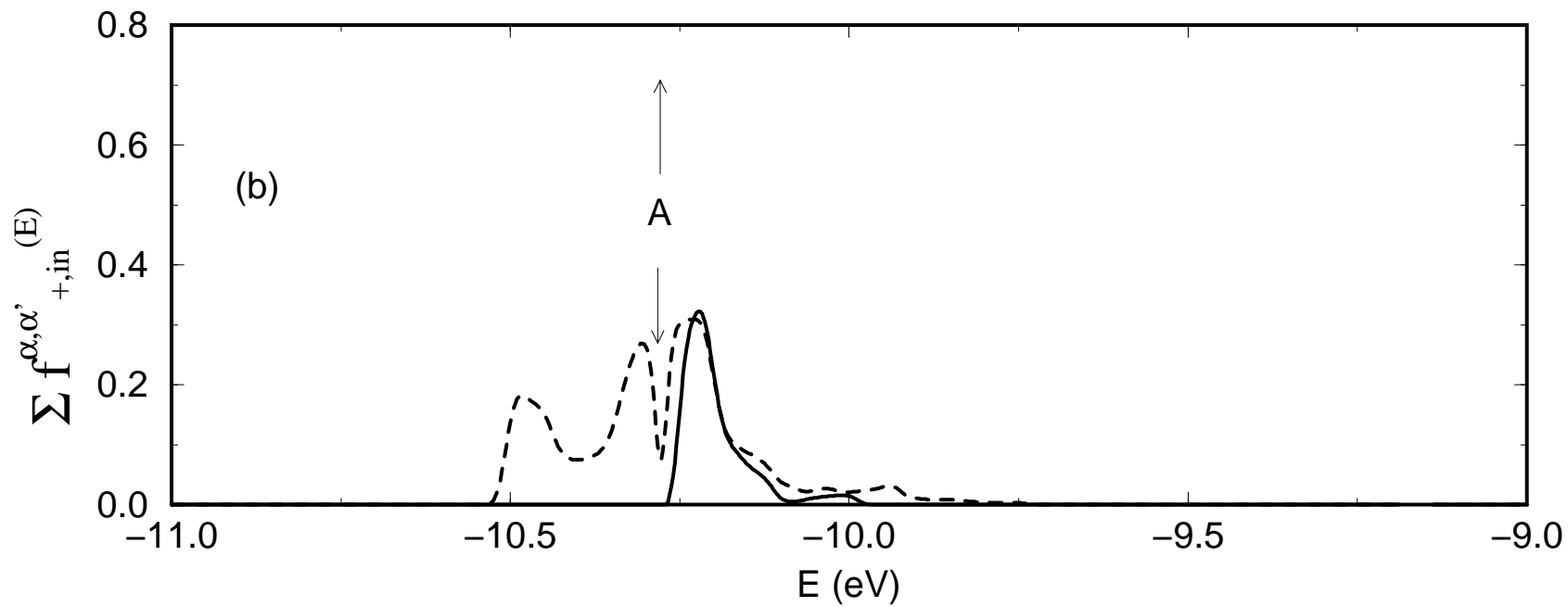
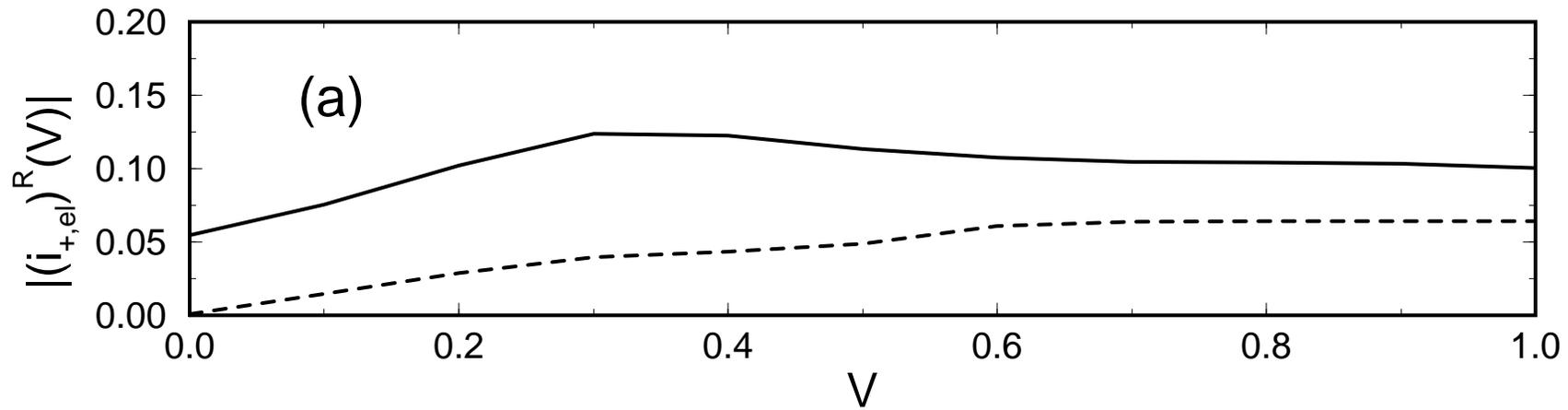
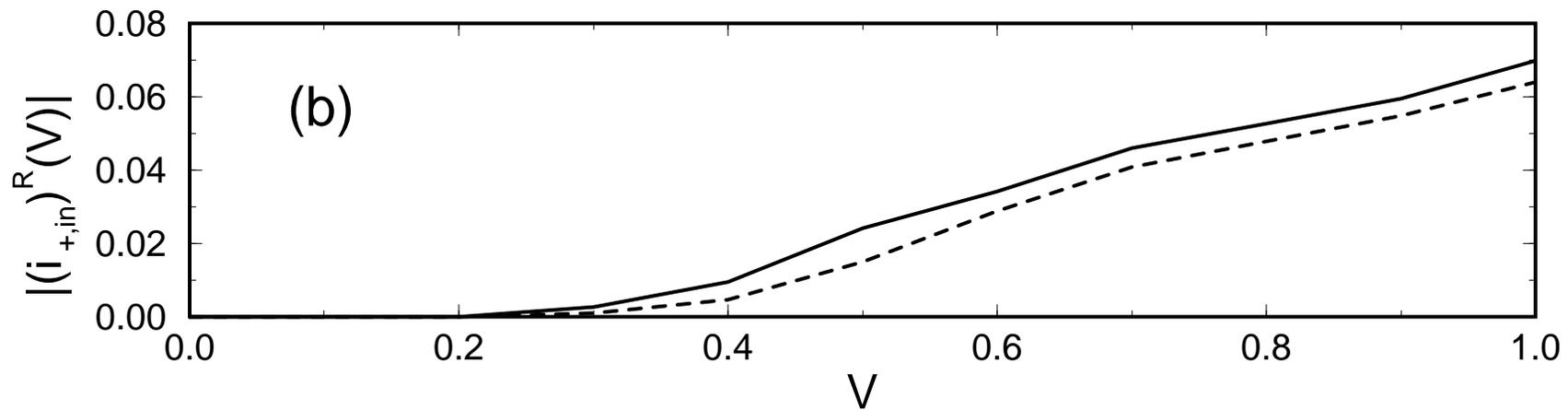
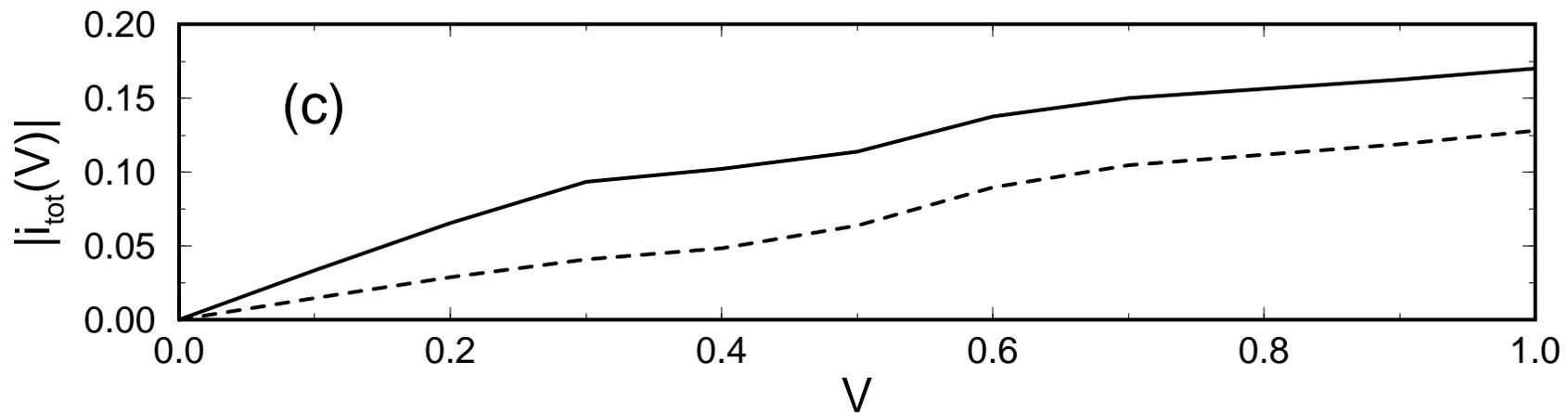
\end{document}